\documentclass[aps,pre,reprint,groupedaddress,showpacs]{revtex4-1}

\usepackage{graphicx}
\usepackage{dcolumn}
\usepackage{bm}

\usepackage{amsmath}
\usepackage{amssymb}
\usepackage{balance}
\usepackage{color}
\usepackage{lastpage}
\usepackage{soul}
\usepackage{hyperref}
\usepackage[utf8]{inputenc}

\renewcommand{\vec}{\bm}

\begin{document}

\preprint{APS/xxxx}

\title{Ternary Free Energy Lattice Boltzmann Model \\ 
with Tunable Surface Tensions and Contact Angles}

\author{Ciro Semprebon$^{a}$, Timm Kr\"uger$^b$ and Halim Kusumaatmaja$^{a}$} 
\email{Email: halim.kusumaatmaja@durham.ac.uk}
\affiliation{$^a$ Department of Physics, Durham University, Durham, DH1 3LE, United Kingdom \\ 
$^{b}$ School of Engineering, The University of Edinburgh, Edinburgh EH9 3JL, United Kingdom}
\date{\today}

\begin{abstract}
We present a new ternary free energy lattice Boltzmann model. The distinguishing feature of 
our model is that we are able to analytically derive and independently vary all fluid-fluid surface tensions and 
the solid surface contact angles. We carry out a number of benchmark tests: 
(i) double emulsions and liquid lenses to validate the surface tensions, (ii) ternary 
fluids in contact with a square well to compare the contact angles against analytical predictions, 
and (iii) ternary phase separation to verify that the multicomponent fluid dynamics is accurately captured. 
Additionally we also describe how the model here presented here can be extended to include an arbitrary 
number of fluid components.
\end{abstract}


\maketitle

\section{Introduction}
\label{sec:intro}

Recently, systems involving three or more fluid phases have attracted considerable interest.
The advent of microfluidics allows us to control emulsions containing mixtures of several immiscible liquids 
\cite{Utada2005, Park2012, Guzowski2015, Haase2014, gao_spreading_2011, luo_deformation_2015}.
Emulsions in return are ubiquitously exploited in the food, pharmaceutical and personal care industries.  
The area of enhanced oil recovery also regularly deals with three or more fluid components (i.e.~water, oil, 
and one or more gaseous phases) \cite{Leverett2013}. 
More recently, there has been growing interest in the so-called liquid 
infused surfaces \cite{Wong2011, Smith2013, Schellenberger2015, wexler_shear-driven_2015} that 
share many advantageous properties of superhydrophobic surfaces with the additional benefit 
of increasing stability for the suspended state. 

The variety of computational approaches developed to solve complex multiphase problems
can generally be divided into two groups,
depending on the width of the fluid-fluid interface: (i) sharp and (ii) diffuse interface models. Our focus in this
paper is on the lattice Boltzmann (LB) method \cite{Chen1998, Succi2001, Aidun2010, Kusumaatmaja2010b} which belongs 
to the latter. Here, the interface spreads over several lattice spacings, and its evolution results from the Navier-Stokes 
and advection-diffusion equations. A key advantage of the diffuse interface models is that the motion of 
the interface does not need to be tracked explicitly. All fluid nodes can be treated on an equal footing 
whether they are in the bulk of the 
fluid or at the interface. As such, diffuse interface models are very convenient for studying problems with complex surface 
geometries, including both chemical and topographical heterogeneities \cite{Kusumaatmaja2006,Kusumaatmaja2007a,Sbragaglia2013, 
Ledesma-Aguilar2011, Harting2008, Connington2013}. The LB algorithm is also highly suitable for parallel 
\cite{williams_optimization_2009} and GPU \cite{schonherr_multi-thread_2011} computing, allowing it to be employed 
in the investigation of flow problems with demanding time and length scales.

While a wide range of LB models have been developed for the binary case \cite{Swift1996,Briant2004a,Inamuro2000,
Zhang2004,Shao2014, shan_lattice_1993, shan_multicomponent_1995,Sbragaglia2007, Gunstensen1991,Grunau1993,Ba2013, 
MazloomiM2015}, systems with three or more fluid components in comparison have received much less attention in the 
LB community. Several ternary models have been proposed to study water-oil-amphiphile mixtures \cite{Chen2000, 
nekovee_lattice-boltzmann_2000,Li2007,Furtado2010},
with two bulk phases, and an amphiphile phase that prefers to locate at the oil-water interface.
There have also been efforts to generalize multi-component LB models to account for an arbitrary number 
of fluid components \cite{Lamura1999,dupin2004,Spencer2010,Leclaire2013a}. 
These works, however, focus solely on bulk behavior, away from solid surfaces. 
As discussed above, for flow problems in enhanced oil recovery, liquid infused surfaces and many others, 
there is a need for a model which allows a thorough control not only of the fluid-fluid surface tensions, 
but also of the fluid-solid interactions.

The key contribution of this paper is to provide a description for a multi-component free energy LB model where it 
is possible to analytically derive and independently control the fluid-fluid surface tensions and the contact angles
that they form with a solid surface. We will focus our discussions on the ternary system, but the model can be 
readily generalized to more fluid components.

The free energy formalism followed in this work is a top-down approach, where we start by writing the free energy of 
the fluids \cite{Swift1996,Briant2004a,Inamuro2000,Zhang2004,Shao2014}. The form of the free energy functional should 
capture the intended features of the thermodynamics of the system, e.g., the miscibility of the components and surface 
tension between different fluids. The corresponding chemical potential, pressure tensor, and LB equation can then be 
subsequently derived from the free energy functional. This is in contrast to the pseudopotential \cite{shan_lattice_1993, 
shan_multicomponent_1995,Sbragaglia2007} and the color \cite{Gunstensen1991,Grunau1993,Ba2013} models that follow a 
bottom-up approach. In a bottom-up model, the starting point is often kinetic theory, and some form of interactions 
are postulated between the fluids at the level of the Boltzmann equation. Reminiscent to many other lattice- and 
particle-based simulation techniques, separation between different fluid phases and components can be induced by 
tuning the interaction potentials. 

The paper is organized as follows. In section~\ref{sec:ThermoDynamics}, we describe the Landau free energy functionals 
that capture the bulk and surface thermodynamics of the ternary fluids. We explicitly derive the predicted values of the 
surface tensions and the contact angles given a small set of input parameters. We also discuss how the model can be extended 
to account for an arbitrary number of fluid components. The continuum equations of motion of the fluids are given in 
section \ref{sec:EoM}. Then, in section \ref{sec:LBM}, we show an LB implementation that captures both the thermodynamics and 
hydrodynamics of the fluids. We provide three sets of benchmark simulations in section \ref{sec:Benchmarks} to validate
various aspects of our model, including (i) double emulsions and liquid lenses, (ii) ternary fluids in contact 
with a square well, and (iii) ternary phase separation. Finally, we summarize the key results of our paper and 
provide an outlook on future work in section \ref{sec:conclusion}.

\section{Thermodynamics}
\label{sec:ThermoDynamics}

We start with a description of the bulk thermodynamics (section~\ref{sec:BulkThermo}), followed by the surface 
thermodynamics (section~\ref{sec:SurfThermo}). We then introduce auxiliary variables in section~\ref{sec:Auxiliary}, 
which allow the model to be implemented easily within the standard LB algorithm.
In section ~\ref{sec:Invert}, we show how the desired contact angles
can be translated to the parameters in our model. We discuss the extension to systems with more than three fluid components
in section~\ref{sec:ExtensionComponents}.

\subsection{Bulk Thermodynamics}
\label{sec:BulkThermo}

One suitable way to define a free energy functional that models three fluid components is given by \cite{Kim2007}
\begin{equation}
 \begin{aligned}
  F &= \int_\Omega \left[\frac{\kappa_1}{2} C_1^2 (1 - C_1)^2 + 
  	\frac{\kappa_2}{2} C_2^2 (1 - C_2)^2 +  \right. \\
  &	\qquad \left. \frac{\kappa_3}{2} C_3^2 (1 - C_3)^2  
  + \frac{\kappa'_1}{2} (\nabla C_1)^2 +  \right. \\
   & \qquad \left.\frac{\kappa'_2}{2} (\nabla C_2)^2 + 
  \frac{\kappa'_3}{2} (\nabla C_3)^2\right]\, \mathrm{d}V
 \end{aligned}
  \label{eq:energy_functional_general}
\end{equation}
where $C_1$, $C_2$ and $C_3$ correspond to the concentration fractions of fluids 1, 2, and 3.
$\Omega$ is the system volume, and the remaining parameters will be discussed below.
By construction, each variable $C_m$ ($m = 1,2,3$) has two bulk minima given by $C_m = 0$ and $1$.
Eq.~\eqref{eq:energy_functional_general} thus, in principle, has $2^3=8$ bulk minimizers. For a ternary fluid system, 
we are not interested in all eight minima, but instead only in the following three:
\begin{equation}
 \begin{aligned}
  C_1 &= 1, \quad C_2 = 0, \quad C_3 = 0; \\
  C_1 &= 0, \quad C_2 = 1, \quad C_3 = 0; \\
  C_1 &= 0, \quad C_2 = 0, \quad C_3 = 1. \\
 \end{aligned}
 \label{eq:bulkminima}
\end{equation}
To strictly ensure we only obtain these minima, one can impose a hard constraint $C_1 + C_2 + C_3 = 1$ 
or introduce an energy penalty term proportional to $(1-C_1-C_2-C_3)^2$.
Here, as we shall see in section~\ref{sec:Auxiliary}, we use a set of variable transformations where the 
(normalized) mass density is defined as $n = C_1 + C_2 + C_3$ and initialized to $n=1$ (in simulation units).
Since our LB algorithm is only weakly compressible (see e.g.~\cite{geller_benchmark_2006}), the density $n$ does not 
deviate far from $C_1 + C_2 + C_3 = 1$, and the additional constraint is not necessary. 

The gradient terms in Eq.~\eqref{eq:energy_functional_general} account 
for the energy penalty for having interfaces. We will now derive how the parameters $\kappa$ 
and $\kappa'$ are related to the interfacial widths and the surface tensions
between the three fluids. Without any loss of generality, let us focus on the interface between 
bulk phases $m$ and $n$ ($m,n =1,2,3$ and $m \neq n$).
We can set the third fluid concentration fraction to zero
everywhere, and exploit the relation $C_n + C_m = 1$ to rewrite Eq.~\eqref{eq:energy_functional_general}
into
\begin{equation}
 \begin{aligned}
  F &= \int_\Omega \left[\frac{\kappa_m}{2} C_m^2 (1 - C_m)^2 + 
  \frac{\kappa_n}{2} C_n^2 (1 - C_n)^2 + \right. \\
  & \qquad  \left.   \frac{\kappa'_m}{2} (\nabla C_m)^2 + 
  \frac{\kappa'_n}{2} (\nabla C_n)^2 \right]\, \mathrm{d}V \\
  &= \int_\Omega \left[\frac{\kappa_m + \kappa_n}{2} C_m^2 (1 - C_m)^2 + \right. \\ 
  & \qquad  \left. \frac{\kappa'_m + \kappa'_n}{2} (\nabla C_m)^2 \right]\, \mathrm{d}V.
 \end{aligned}
 \label{eq:interfacemn}
\end{equation}

We notice that the simplified free energy in Eq.~\eqref{eq:interfacemn}
has the same structure as for the binary fluid problem \cite{Briant2004a,Kusumaatmaja2010b}. 
As such, we can proceed in the same way. We can define the chemical potential for component $m$ as
\begin{eqnarray}
 \mu_m &=& \frac{\delta F}{\delta C_m} \\
 &=& (\kappa_m + \kappa_n)\left(2 C_m^3 - 3 C_m^2 
 + C_m - \frac{\kappa'_m + \kappa'_n}{\kappa_m 
 + \kappa_n}\nabla^2 C_m\right).  \nonumber
\end{eqnarray}

At \emph{thermodynamic} equilibrium we have $ \mu_m = 0$. Assuming the interface is located at $x = 0$, 
the interfacial profile along the $x$-axis for the concentration of component $m$ is
\begin{equation}
 C_m = \frac{1 + \tanh \frac{x}{2\alpha}}{2} \label{tanh}
\end{equation}
where the parameter $\alpha = \sqrt{(\kappa'_m + \kappa'_n)/(\kappa_m + \kappa_n)}$ 
is proportional to the interface width. It is easy to verify that $C_m \rightarrow 1$ for $x \rightarrow \infty$, 
and $C_m \rightarrow 0$ for $x \rightarrow -\infty$. To obtain the surface tension $\gamma_{mn}$, 
we substitute the concentration profile in Eq.~\eqref{tanh} 
into Eq.~\eqref{eq:interfacemn} and compute the excess free energy per unit area:
\begin{equation}
 \begin{aligned}
  \gamma_{mn} &= \int_{-\infty}^{+\infty} 
  \left[\frac{\kappa_m + \kappa_n}{2} C_m^2 (1 - C_m)^2 + \right. \\ 
  & \qquad  \left. \frac{\kappa'_m + \kappa'_n}{2} (\nabla C_m)^2 \right]\, \mathrm{d}x \\
     &=\frac{\sqrt{(\kappa'_m + \kappa'_n)(\kappa_m + \kappa_n)}}{6}. \label{eq:SurfTensGen}
 \end{aligned}
\end{equation}

The parameters $\kappa$ and $\kappa'$ can be arbitrarily tuned to achieve the desired surface 
tensions (e.g., to reproduce experimental parameters). However, there are two constraints: $\kappa_m > 0$ 
and $\kappa'_m + \kappa'_n > 0$. The former is needed to ensure that we have two coexisting minima at 
$C_m = 0$ and $C_m = 1$ for every concentration fraction. The latter is required to have 
positive surface tensions. This constraint can be relaxed if negative surface 
tensions are indeed a desired feature in the simulations. 

For most applications, it is convenient to reduce the number of free parameters in the 
$\{\kappa,\kappa'\}$ space since the extra degrees of freedom are not always required. For the rest of the 
paper, we will set $\kappa' = \alpha^2 \kappa$ for all components. This simplification ensures that the 
interface width is the same for all three fluid-fluid interfaces and can be tuned by varying
the value of $\alpha$. In this case, the formula for the surface tension becomes
\begin{equation}
 \begin{aligned}
  \gamma_{mn} =\frac{\alpha}{6}(\kappa_m + \kappa_n). \label{eq:SurfTens}
 \end{aligned}
\end{equation}
It is worth noting that in this model, if $\kappa_l$, $\kappa_m$, $\kappa_n>0$ and $l \not= m \not= n \not= l$, 
the following relation is always true: $\gamma_{lm} + \gamma_{mn} > \gamma_{ln}$.

\subsection{Surface Thermodynamics}
\label{sec:SurfThermo}

The different affinities of fluids to the solid surface are usually quantified by the material property 
named \emph{contact angle}. 
If the subscripts $m,n$ denote the bulk fluid phases and $s$ the solid surface, then the contact 
angle with respect to the fluid phase $m$ is given by \cite{Bonn2009a}
\begin{equation} 
	\cos{\theta_{mn}} = \frac{\gamma_{sn}-\gamma_{sm}}{\gamma_{mn}}. 
	\label{eq:contact angle}
\end{equation}
Here $\gamma_{sm}$, $\gamma_{sn}$ and 
$\gamma_{mn}$ are, respectively, the surface tensions between the solid and fluid phase $m$, the solid and fluid 
phase $n$ and the two fluid phases. 

We will now show how the wetting boundary conditions are implemented in our model and how the
contact angles can be tuned simultaneously by introducing a small number of parameters. 
We will once again follow a thermodynamic route and describe the surface free energy 
contributions by
\begin{equation}
  \int_{\partial\Omega} \left[\Psi_1 \rvert_s + \Psi_2 \rvert_s + \Psi_3 \rvert_s \right]\, \mathrm{d}S. \label{eq:surfaceenergy}
\end{equation}
Following Cahn \cite{Cahn1977}, the surface free energy density for each component can be expressed as
\begin{equation}
  \Psi_m \rvert_s =-h_m C_m \rvert_s \label{eq:Cahn}
\end{equation}
where $C_m \rvert_s$ is the value of the order parameter $C_m$ at the solid boundary, and the parameter $h_m$ still has to be specified.
Employing standard tools of calculus of variation, functional minimisation for the component $m$ 
at the solid boundary leads to the condition
 \begin{equation}
 \alpha^2 \kappa_m \nabla_\perp C_m \rvert_s = \left. \frac{\mathrm{d} \Psi_m}{\mathrm{d} C_m} \right\rvert_s = -h_m.
 \label{eq:cahn_gradient}
\end{equation}
Here $\nabla_\perp$ defines the perpendicular derivative of the concentration with respect to the solid surface.
We can also take advantage of Noether's theorem to show that 
\begin{equation}
 \begin{aligned}
 \frac{\kappa_m}{2} C_m^2 (1 - C_m)^2 
 - \alpha^2\frac{\kappa_m}{2} (\nabla C_m)^2 = \mathrm{const} = 0.
 \end{aligned}
  \label{eq:invariant}
\end{equation}
By evaluating the expression on the left-hand side far from the interface, we can conclude that
the constant value on the right-hand side is zero.

Let us now compute the surface tension between the solid surface and the fluid component $m$, $\gamma_{sm}$.
It is worth noting here that the contributions to this surface tension come not only from the majority phase $m$,
but also from the other two minority phases, $l,n \neq m$. Furthermore, in addition to the term in 
Eq.~\eqref{eq:surfaceenergy}, we also have to account for the variation of the concentration fractions from 
their bulk values in Eq.~\eqref{eq:energy_functional_general}
to properly account for the fluid-solid surface tensions.

We will first focus on the contribution $I_m$ to the surface tension $\gamma_{sm}$ from the majority phase $m$. 
If $x = 0$ is the location of the fluid-solid interface and $x > 0$ is the fluid region, this is given by
\begin{eqnarray}
 I_m &=& -h_m C_m \rvert_s \nonumber \\
 &&+ \int_0^\infty  \frac{\kappa_m}{2} \left[ C_m^2 (1 - C_m)^2 +
 \alpha^2\left(\frac{\mathrm{d} C_m}{\mathrm{d} x}\right)^2\right] \mathrm{d}x \nonumber \\
 &=& -h_m C_m \rvert_s + \int_0^\infty \alpha^2 \kappa_m \left(\frac{\mathrm{d} C_m}{\mathrm{d} x}\right)^2 \mathrm{d}x.
 \label{eq:surf_energy_component}
\end{eqnarray}
The integral accounts for the contribution in the transition region where $C_m$ varies between the values
at the boundary and in the bulk. We have also employed Noether's theorem to simplify the integral term.

The value of $C_m \rvert_s$ can be determined as follows. 
From the invariant condition in Eq.~\eqref{eq:invariant}, we can write
\begin{equation}
  \left(\frac{\mathrm{d} C_m}{\mathrm{d}x}\right)^2 = \frac{1}{\alpha^2} C_m^2 (1 - C_m)^2.
 \label{eq:gradient_replace}
 \end{equation}
Substituting Eq.~\eqref{eq:gradient_replace} into Eq.~\eqref{eq:cahn_gradient}, we further obtain
\begin{equation}
 \alpha^2\kappa_m^2 (C_m \rvert_s)^2 (1 - C_m \rvert_s)^2 = h_m^2.
 \label{eq:bulk_free_energy}
\end{equation}
In general, Eq.~\eqref{eq:bulk_free_energy} has four solutions
\begin{equation}
 C_m \rvert_s(\kappa_m,h_m)=\frac{1}{2}\left(1 \pm \sqrt{1 \pm \frac{4 h_m}{\alpha \kappa_m}}  \right).
\end{equation}
However, only two of them are physical. To elucidate
this statement, let us assume $h_m > 0$. As such, the fluid $m$ has an attractive interaction
with the surface. We expect the concentration of fluid $m$ to be larger near the surface compared to 
its bulk value. The appropriate solutions are therefore
\begin{eqnarray}
 C_m^\mathrm{max} \rvert_s (\kappa_m,h_m)=\frac{1}{2}\left(1 + \sqrt{1 + \frac{4 h_m}{\alpha \kappa_m}}  \right), \\
 C_m^\mathrm{min} \rvert_s (\kappa_m,h_m)=\frac{1}{2}\left(1 - \sqrt{1 - \frac{4 h_m}{\alpha \kappa_m}}  \right).
\end{eqnarray}
The former is the suitable solution when $C_m$ is the majority phase (i.e., $C_m=1$ in the bulk), 
the latter when $C_m$ is a minority phase (i.e., $C_n=0$ in the bulk).

Given $C_m \rvert_s$, the integral in Eq.~\eqref{eq:surf_energy_component} can be computed as 
\begin{eqnarray}
 && \int_0^\infty\alpha^2\kappa_m\left(\frac{\mathrm{d} C_m}{\mathrm{d} x}\right)^2 \mathrm{d}x  = \int_{C_m^\mathrm{max} \rvert_s }^{1} \alpha\kappa_m C_m (1-C_m)\, \mathrm{d}C_m \nonumber \\
 && = \frac{\alpha}{12}\left(\kappa_m + \sqrt{1+\frac{4 h_m}{\alpha\kappa_m}}
 \left(\frac{2 h_m}{\alpha} -\kappa_m\right) \right). 
 \label{eq:integral_max}
\end{eqnarray}

Adopting the same analysis leading to Eq.~\eqref{eq:surf_energy_component}, the contributions 
$J_n$ to $\gamma_{sm}$ from the minority phases $n \neq m$ can be readily derived:
\begin{eqnarray}
 J_n & = &-h_n C_n \rvert_s + \int_0^\infty  \kappa_n \alpha^2\left(\frac{\mathrm{d} C_n}{\mathrm{d} x}\right)^2 \mathrm{d}x \nonumber \\
 & =& -h_n C_n \rvert_s - \int_{C_m^\mathrm{min} \rvert_s}^{0} \alpha\kappa_n C_n (1-C_n)\, \mathrm{d}C_n \\
 & =& -h_n C_n \rvert_s +\frac{\alpha}{12}\left(\kappa_n - \sqrt{1-\frac{4 h_n}{\alpha\kappa_n}}
   \left(\frac{2 h_n}{\alpha} +\kappa_n\right) \right). \nonumber
\end{eqnarray}
Thus, summing the contributions from the majority and minority phases, the interfacial tension $\gamma_{sm}$ is given by
\begin{equation}
\gamma_{sm} = I_m +  \sum_{n\neq m} J_n. \label{eq:solidST}
\end{equation}
This relation is valid for an arbitrary number of phases. Most importantly, the results from this subsection show that:
(i) the parameters $\{h_1,h_2,h_3\}$ enter the simulations through the Neumann boundary condition in Eq.~\eqref{eq:cahn_gradient};
and (ii) given the parameters $\{h_1,h_2,h_3\}$, the solid-fluid surface tensions $\{\gamma_{s1},\gamma_{s2},\gamma_{s3}\}$
and subsequently the contact angles $\{\theta_{12},\theta_{23},\theta_{31}\}$ can be computed analytically. 
The latter follow from Young's equation, Eq. \eqref{eq:contact angle}.

\subsection{Auxiliary variables}
\label{sec:Auxiliary}

So far, the description of the thermodynamics of the ternary fluid is carried out in terms of the concentration
fractions. As discussed above, to ensure the condition $C_1+C_2+C_3=1$, one can introduce an additional
term proportional to $(1-C_1-C_2-C_3)^2$ in the free energy functional. This additional term, however,
would complicate the derivations of the thermodynamic quantities.  A more elegant solution involves
a variable transformation, defining three auxiliary fields $\rho$, $\phi$ and $\psi$:
\begin{equation}
 \rho = C_1+C_2+C_3, \quad  \phi = C_1 - C_2, \quad \psi = C_3.
\end{equation}
Additionally we ensure that the (dimensionless) mass density is initialized to $\rho = 1$. 
Here we have assumed that all fluid components have the same density. 
Another key advantage of the variable transformation is that the continuity and Navier-Stokes equations 
are generally written in terms of $\rho$ rather than $C_1,C_2$ and $C_3$. 
The original fields can be expressed in terms of the new fields as
$C_1 = (\rho + \phi - \psi)/2$, $C_2 = (\rho - \phi - \psi)/2$, and $C_3 = \psi$.

Substituting the new variables, the free energy functional assumes the form
\begin{equation}
 \begin{aligned}
  F & = \int_\Omega \Bigl[\frac{\kappa_1}{32} (\rho + \phi - \psi)^2 (2 + \psi - \rho- \phi)^2 \\
  & \quad \left. + \frac{\alpha^2\kappa_1}{8} \left(\nabla \rho + \nabla \phi - \nabla \psi\right)^2 \right. \\
  & \quad + \frac{\kappa_2}{32} (\rho - \phi - \psi)^2 (2 + \psi -\rho + \phi)^2 \\
  & \quad + \frac{\alpha^2\kappa_2}{8} \left(\nabla \rho - \nabla \phi - \nabla \psi\right)^2 \\
  & \quad + \frac{\kappa_3}{2} \psi^2 (1 - \psi)^2 + \frac{\alpha^2\kappa_3}{2} (\nabla \psi)^2 \Bigr]\, \mathrm{d}V.
 \end{aligned}
 \label{eq:functional_newvar}
\end{equation}

The results we have derived in the previous sections require little to no changes upon 
these variable transformations. The fluid-fluid surface tension is still given by Eq.~\eqref{eq:SurfTens}.
The wetting boundary conditions now need to be implemented on the variables $\{\rho, \phi, \psi\}$ 
rather than  $\{C_1, C_2, C_3\}$ such that
\begin{eqnarray}
&&\nabla_\perp \rho \rvert_s = -\frac{h_1}{\alpha^2\kappa_1} - \frac{h_2}{\alpha^2\kappa_2} - \frac{h_3}{\alpha^2\kappa_3} , \\
&&\nabla_\perp \phi \rvert_s = -\frac{h_1}{\alpha^2\kappa_1} + \frac{h_2}{\alpha^2\kappa_2} , \\
&&\nabla_\perp \psi \rvert_s = -\frac{h_3}{\alpha^2\kappa_3} .
\end{eqnarray}

\subsection{Inverting the contact angle relations}
\label{sec:Invert}

In section \ref{sec:SurfThermo}, given the parameters $h_1$, $h_2$ and $h_3$, we discussed 
how the contact angles of the fluids can be derived.
Here we will describe how to invert this relation.

First, we note that in presence of a homogeneous substrate not all three contact angles 
are actually independent. The contact angles in terms of the surface tensions are
\begin{eqnarray}
\gamma_{12} \cos \theta_{12} = \gamma_{2s}-\gamma_{1s}, \\
\gamma_{23} \cos \theta_{23} = \gamma_{3s}-\gamma_{2s}, \\
\gamma_{31} \cos \theta_{31} = \gamma_{1s}-\gamma_{3s} .
\end{eqnarray} 
Summing these three equations, we obtain what is often called as the \emph{Girifalco-Good relation} \cite{Girifalco1957}:
\begin{equation}
\label{eq:GoodGirifalco}
\gamma_{12} \cos \theta_{12} + \gamma_{23} \cos \theta_{23} + \gamma_{31} \cos \theta_{31} = 0 .
\end{equation} 

To invert the contact angle relations, we can derive how the contact angle $\theta_{mn}$ depends on $h_m$ and $h_n$,
 as well as on $\kappa_m$ and $\kappa_n$:
\begin{eqnarray}
\cos \theta_{mn} &=& \frac{(\alpha \kappa_n + 4h_n)^{3/2} - 
(\alpha \kappa_n - 4h_n)^{3/2}}{2(\kappa_m+\kappa_n) (\alpha\kappa_n)^{1/2}}
\label{eq:invertCA} \\
&-& \frac{(\alpha \kappa_m + 4h_m)^{3/2} - (\alpha \kappa_m - 4h_m)^{3/2}}{2(\kappa_m+\kappa_n) (\alpha\kappa_m)^{1/2}}. \nonumber
\end{eqnarray} 
In practice, $\kappa_m$ and $\kappa_n$ are determined by our choice of the fluid-fluid surface tensions.
Due to the aforementioned Girifalco-Good relation, only two out of the three contact angles are independent; 
yet we have introduced three parameters $h_1$, $h_2$ and $h_3$.
This implies that there is an infinite set of $h$ parameters able to reproduce
a given combination of contact angles ($\theta_{12}$, $\theta_{23}$ and $\theta_{31}$).

There are several options to remove the redundancy in the $h$ parameters.
In our simulations, we usually require the gradient of the density to be zero at the surface, 
$\nabla_\perp \rho \rvert_s = 0$, such that
\begin{equation}
\frac{h_1}{\kappa_1} + \frac{h_2}{\kappa_2} + \frac{h_3}{\kappa_3} = 0. \label{eq:Redundancy}
\end{equation}
Combining Eq.~\eqref{eq:invertCA} and Eq.~\eqref{eq:Redundancy}, we can uniquely determine $h_1$, $h_2$ 
and $h_3$ given a prescribed set of contact angles.

Let us now comment on the physical meaning of this redundancy in the $h$ parameters. Our thermodynamic 
model allows a one-to-one mapping between $h_1$, $h_2$ and $h_3$ and the fluid-solid surface tensions 
$\gamma_{1s}$, $\gamma_{2s}$ and $\gamma_{3s}$. However, for the computation of the contact angles, 
only \emph{differences} in the fluid-solid surface tensions are important. Setting the condition in 
Eq.~\eqref{eq:Redundancy} is equivalent to adding/removing a given constant to all the fluid-solid 
surface tensions. The advantage of imposing 
Eq.~\eqref{eq:Redundancy} is that the mass density is not affected by surface forces and remains 
close to $\rho = 1$ throughout the simulation domain.

\subsection{Extension to more than three fluid components}
\label{sec:ExtensionComponents}

The model proposed here can be generalized to include an arbitrary number of fluid components.
For $N>3$ bulk fluids, a suitable Landau free energy functional is
\begin{equation}
 \begin{aligned}
  F = &\sum_{m=1}^N  \int_\Omega \left[\frac{\kappa_m}{2} C_m^2 (1 - C_m)^2 + 
   \frac{\alpha^2 \kappa_m}{2} (\nabla C_m)^2 \right] \, \mathrm{d}V \\
    & +\sum_{m=1}^N  \int_{\partial\Omega} -h_m C_m \, \mathrm{d}S.
 \end{aligned}
\end{equation}
The derivations of the fluid-fluid and fluid-solid surface tensions follow exactly the same routes as
those leading to Eq.~\eqref{eq:SurfTens} and Eq.~\eqref{eq:solidST}. 
We may also introduce a similar form of variable transformations where 
$\rho = \sum_{m=1}^N C_m$, $\phi = C_1-C_2$, and $\psi_l = C_l$ with $l>2$.
In this case, the wetting boundary conditions are given by
\begin{equation}
\begin{aligned}
 \nabla_\perp \rho \rvert_s &= \sum_{m=1}^N -\frac{h_m}{\alpha^2\kappa_m}, \\
 \nabla_\perp \phi \rvert_s &= -\frac{h_1}{\alpha^2\kappa_1} + \frac{h_2}{\alpha^2\kappa_2} , \\
 \nabla_\perp \psi_l \rvert_s &= - \frac{h_l}{\alpha^2\kappa_l} .
\end{aligned}
\end{equation}
Following the arguments in section \ref{sec:Invert}, we emphasize again that not all contact angles
are independent. For $N$ fluids, there are only $N(N-1)/2 -1$ independent contact angles.
In this case, the generalized Girifalco-Good relation reads
\begin{equation}
\sum_{m, n \neq m} \gamma_{mn} \cos \theta_{mn} = 0.
\end{equation}

\section{Equations of Motion}
\label{sec:EoM}

Before we write down the LB equations for a ternary fluid system, let us review the corresponding continuum equations of motion. 
The fluid motion is described by the continuity and Navier-Stokes equations:
\begin{eqnarray} 
&\partial_t \rho + \partial_\gamma \left( \rho v_\gamma \right) = 0 , \label{eq:Continuity} \\
& \partial_t (\rho v_\alpha) + \partial_\beta \left( \rho v_\alpha v_\beta \right) = 
- \partial_\alpha p + \partial_\beta \eta \left( \partial_\beta v_\alpha + \partial_\alpha v_\beta \right) \nonumber \\
& -\rho \partial_\alpha \mu_\rho - \phi \partial_\alpha \mu_\phi  - \psi \partial_\alpha \mu_\psi.
\label{eq:NavierStokes}
\end{eqnarray}
Here, $\vec{v}$ is the fluid velocity, $p$ is the isotropic pressure (usually taken to be the 
ideal gas pressure in the LB method, $p = \rho c_\mathrm{s}^2$ \cite{hou_simulation_1995}), and $\eta$ is 
the dynamic viscosity of the fluid that may depend on the order parameters.

The key additional physics due to the thermodynamics of the ternary fluid is contained in the 
last three terms of Eq.~\eqref{eq:NavierStokes}, corresponding to the three auxiliary fields we 
have introduced in our model. In \emph{mechanical} equilibrium, the chemical potential has to be the same everywhere. 
Any inhomogeneity leads to a body force proportional to the gradient of the chemical potential. 
The chemical potentials corresponding to the distributions of $\rho$, $\phi$, and $\psi$ are
\begin{eqnarray}
  \mu_\rho &=& \frac{\delta F}{\delta \rho} \\
  & = & \frac{\kappa_1}{8} (\rho + \phi-\psi)(\rho + \phi -\psi-2)(\rho + \phi - \psi-1) \nonumber  \\
  &-&  \frac{\kappa_2}{8} (\rho-\phi-\psi)(\rho-\phi - \psi-2)(\rho-\phi-\psi-1)\nonumber  \\
  &+&  \frac{\alpha^2}{4}\left[ (\kappa_1+\kappa_2)( \nabla^2 \psi - \nabla^2 \phi) +(\kappa_2-\kappa_1) \nabla^2 \rho \right], \nonumber
\end{eqnarray}
\begin{eqnarray}  
 \mu_\phi &=& \frac{\delta F}{\delta \phi} \\
  & =& \frac{\kappa_1}{8} (\rho + \phi-\psi)(\rho + \phi -\psi-2)(\rho + \phi - \psi-1) \nonumber  \\
  &-& \frac{\kappa_2}{8} (\rho-\phi-\psi)(\rho-\phi - \psi-2)(\rho-\phi-\psi-1) \nonumber \\
  &+& \frac{\alpha^2}{4}\left[ (\kappa_2-\kappa_1)( \nabla^2 \rho - \nabla^2 \psi) - (\kappa_1+\kappa_2) \nabla^2 \phi \right], \nonumber  
\end{eqnarray}
\begin{eqnarray}
\mu_\psi &=& \frac{\delta F}{\delta \psi} \\
  &=& -\frac{\kappa_1}{8} (\rho + \phi-\psi)(\rho + \phi -\psi-2)(\rho + \phi - \psi-1) \nonumber  \\
  &-& \frac{\kappa_2}{8} (\rho-\phi-\psi)(\rho-\phi - \psi-2)(\rho-\phi-\psi-1) \nonumber  \\
  &+& \kappa_3 \psi (\psi-1)(2\psi-1) + \frac{\alpha^2}{4}\left[ (\kappa_1+\kappa_2) \nabla^2 \rho  \right. \nonumber  \\
  &-& \left.	(\kappa_2-\kappa_1) \nabla^2 \phi - (\kappa_2+\kappa_1+4\kappa_3) \nabla^2 \psi \right]. \nonumber 
\end{eqnarray}

These thermodynamic terms can be implemented in two different approaches in the LB algorithm. 
First, we can apply a body force by employing standard forcing  methods 
(e.g., Guo \cite{guo_discrete_2002} or Shan-Chen \cite{shan_lattice_1993} forcing). Secondly, the thermodynamic terms 
can be taken into account within the definition of the generalized pressure tensor. This second approach
is the one utilized in the present work. In this case, the pressure tensor satisfies the condition 
\begin{equation}
	\partial_\beta P_{\alpha\beta} = \partial_\alpha p + \rho \partial_\alpha \mu_\rho 
	+ \phi \partial_\alpha \mu_\phi + \psi \partial_\alpha \mu_\psi
\end{equation}
and reads
\begin{eqnarray}
 &&P_{\alpha \beta} =   p_\mathrm{b}\delta_{\alpha\beta} \\
 && +\alpha^2 \kappa_{\rho\rho} \left[ (\partial_\alpha \rho)(\partial_\beta \rho) 
   -(1/2)(\partial_\gamma \rho)^2\delta_{\alpha\beta} -\rho (\partial_{\gamma\gamma} \rho)\delta_{\alpha\beta}   \right] \nonumber \\
 && +\alpha^2 \kappa_{\phi\phi} \left[ (\partial_\alpha \phi)(\partial_\beta \phi ) 
   -(1/2)(\partial_\gamma \phi)^2\delta_{\alpha\beta} -\phi (\partial_{\gamma\gamma} \phi)\delta_{\alpha\beta}   \right] \nonumber \\
 && +\alpha^2 \kappa_{\psi\psi} \left[ (\partial_\alpha \psi)(\partial_\beta \psi ) 
   -(1/2)(\partial_\gamma \psi)^2\delta_{\alpha\beta} -\psi (\partial_{\gamma\gamma} \psi)\delta_{\alpha\beta}   \right] \nonumber \\
 && +\alpha^2 \kappa_{\rho\phi} \left[ (\partial_\alpha \rho)(\partial_\beta \phi) +(\partial_\alpha \phi)(\partial_\beta \rho) \right. \nonumber \\
 && \qquad \left.  -(\partial_\gamma \rho)(\partial_\gamma \phi)\delta_{\alpha\beta} 
   -\rho (\partial_{\gamma\gamma} \phi)\delta_{\alpha\beta} -\phi (\partial_{\gamma\gamma} \rho)\delta_{\alpha\beta}  \right] \nonumber \\
 && +\alpha^2 \kappa_{\rho\psi} \left[ (\partial_\alpha \rho)(\partial_\beta \psi) +(\partial_\alpha \psi)(\partial_\beta \rho) \right. \nonumber \\
 && \qquad \left.  -(\partial_\gamma \rho)(\partial_\gamma \psi)\delta_{\alpha\beta} 
   -\rho (\partial_{\gamma\gamma} \psi)\delta_{\alpha\beta} -\psi (\partial_{\gamma\gamma} \rho)\delta_{\alpha\beta}  \right] \nonumber \\
 && +\alpha^2 \kappa_{\phi\psi} \left[ (\partial_\alpha \phi)(\partial_\beta \psi) +(\partial_\alpha \psi)(\partial_\beta \phi) \right. \nonumber \\
 && \qquad \left. -(\partial_\gamma \phi)(\partial_\gamma \psi)\delta_{\alpha\beta} 
   -\phi (\partial_{\gamma\gamma} \psi)\delta_{\alpha\beta} -\psi (\partial_{\gamma\gamma} \phi)\delta_{\alpha\beta}  \right] \nonumber
\end{eqnarray}
where the mixing coefficients can be derived by collecting the appropriate gradient terms:
\begin{equation}
 \begin{aligned}
 & \kappa_{\rho\rho} = \kappa_{\phi\phi} = \frac{\kappa_1+\kappa_2}{4} , 
 & \kappa_{\psi\psi} =\frac{\kappa_1+\kappa_2+4\kappa_3}{4},  \\
 & \kappa_{\rho\phi} = -\kappa_{\phi\psi} =\frac{\kappa_1-\kappa_2}{4}, 
 & \kappa_{\rho\psi} =-\frac{\kappa_1+\kappa_2}{4}. \nonumber
 \end{aligned}
\end{equation}
The bulk pressure term $p_\mathrm{b}$ is given by
\begin{eqnarray}
 && p_\mathrm{b}  =  \rho c_\mathrm{s}^2  + (\kappa_1+\kappa_2 ) \left[ \frac{3}{32} \rho^4+ \frac{3}{32} \phi^4 + \frac{9}{16} \rho^2\phi^2  \right.  \\
 && + \frac{9}{16} \rho^2 \psi^2 + \frac{9}{16} \phi^2 \psi^2 - \frac{3}{8} \rho^3\psi - \frac{3}{8} \rho \psi^3 + \frac{3}{4} \rho^2\psi -\frac{3}{4}\rho\phi^2 \nonumber \\
 && \left. -\frac{3}{4}\rho\psi^2+\frac{3}{4}\phi^2\psi - \frac{1}{4}\rho^3 + \frac{1}{8} \rho^2+ \frac{1}{8} \phi^2 - \frac{1}{4}\rho\psi - \frac{9}{8}\rho\phi^2\psi \right] \nonumber \\
  && + (\kappa_1-\kappa_2) \left[ \frac{3}{8} \rho^3\phi+\frac{3}{8}\rho\phi^3-\frac{3}{8}\phi^3\psi-\frac{3}{8} \phi\psi^3 - \frac{1}{4} \phi^3 \right. \nonumber \\
 && - \frac{3}{4} \rho^2\phi - \frac{3}{4} \phi\psi^2 + \frac{1}{4} \rho\phi - \frac{1}{4} \phi\psi
 + \frac{9}{8} \rho\phi\psi^2 - \frac{9}{8} \rho^2\phi\psi \nonumber \\
 && \left. + \frac{3}{2} \rho\phi\psi \right]  + \frac{1}{4}(\kappa_1+\kappa_2-8\kappa_3)\psi^3
  + (\kappa_1+\kappa_2+ 16 \kappa_3 ) \nonumber \\
  && \left[\frac{3}{32}  \psi^4 + \frac{1}{8}\psi^2\right]. \nonumber
 \end{eqnarray}

The order parameters $\phi$ and $\psi$ themselves evolve through advection-diffusion (Cahn-Hilliard) equations:
\begin{eqnarray} 
        \partial_t \phi  + \partial_\alpha (\phi v_\alpha)  = M_\phi \nabla^2 \mu_\phi ,\label{eq:Cahn-Hilliard1} \\
        \partial_t \psi  + \partial_\alpha (\psi v_\alpha)  = M_\psi \nabla^2 \mu_\psi . \label{eq:Cahn-Hilliard2}
\end{eqnarray}
The second term on the left-hand side is the advection term. The diffusive term on the right-hand side accounts 
for motion of the order parameter due to inhomogeneities in the chemical potential. In principle the mobility 
parameters $M_\phi$ and $M_\psi$ can be inhomogeneous and varied independently. However, of particular use is 
the special case where the original fields $C_1$, $C_2$, and $C_3$ have identical mobility parameters. In our 
current notation, this is achieved by setting  $M_\phi = 3 M_\psi$ (see the derivation in appendix~\ref{app:MobilityParameters}). 

\section{The Lattice Boltzmann Equation}
\label{sec:LBM}

We now describe an LB algorithm that solves Eq.~\eqref{eq:Continuity}, Eq.~\eqref{eq:NavierStokes}, 
Eq.~\eqref{eq:Cahn-Hilliard1}, and Eq.~\eqref{eq:Cahn-Hilliard2}. For a ternary fluid system, 
we need to define three distribution functions, $f_i({\bf r},t)$, $g_i({\bf r}, t)$ and $k_i({\bf r}, t)$, 
corresponding to the density of the fluid $\rho$ and the two order parameters $\phi$ and $\psi$. 
The physical variables are then related to the distribution functions by \cite{Succi2001}
\begin{eqnarray}
 & \rho({\bf r}, t) = \sum_i f_i({\bf r}, t),  \;\;\;  & \rho({\bf r}, t)v_{\alpha}({\bf r}, t) = \sum_i f_i({\bf r}, t) e_{i\alpha} , \nonumber \\
 & \phi({\bf r}, t) = \sum_i g_i({\bf r}, t),  \;\;\;  & \psi({\bf r}, t) = \sum_i k_i({\bf r}, t) . \label{lbconstraint}
\end{eqnarray}
The quantities $e_{i\alpha}$ correspond to the standard lattice velocities in the LB method.
$\Delta x$ and $\Delta t$ are the lattice spacing and time step respectively.
Here we implement the D3Q19 model with 19 velocities in three dimensions for which we have 
$\vec{e}_{i} = \Delta x/ \Delta t$ $\{(0,0,0)$, $(\pm1,0,0)$, 
$(0,\pm1,0)$, $(0,0,\pm1)$, $(\pm1,\pm1,0)$, $(\pm1,0,\pm1)$, $(0,\pm1,\pm1) \}$.

For the sake of clarity, we describe the LB implementation using a standard BGK single-relaxation-time approach. 
The extension to multiple relaxation times is straightforward, similar to that described by Pooley et al. 
\cite{Pooley2008} for the binary free energy model. The collision step is given by
\begin{eqnarray}
&&f_i^\star({\bf r}, t) = f_i({\bf r}, t) - \tfrac{\Delta t }{\tau} \left[f_i({\bf r}, t) - f_i^{\mathrm{eq}}({\bf r}, t)  
\right], \label{latboltCollision} \nonumber \\
&&g_i^\star({\bf r}, t) = g_i({\bf r}, t) - \tfrac{\Delta t }{\tau_\phi} \left[g_i({\bf r}, t) - g_i^{\mathrm{eq}}({\bf r}, t) \right], \\
&&k_i^\star({\bf r}, t) = k_i({\bf r}, t) - \tfrac{\Delta t }{\tau_\psi} \left[k_i({\bf r}, t) - k_i^{\mathrm{eq}}({\bf r}, t) \right]. \nonumber 
\end{eqnarray}
The propagation step reads
\begin{eqnarray}
&&f_i({\bf r} + {\bf e}_i \Delta t , t+\Delta t) = f_i^\star({\bf r}, t), \nonumber \\
&&g_i({\bf r} + {\bf e}_i \Delta t , t+\Delta t) = g_i^\star({\bf r}, t), \label{latboltPropagation} \\
&&k_i({\bf r} + {\bf e}_i \Delta t , t+\Delta t) = k_i^\star({\bf r}, t). \nonumber 
\end{eqnarray}
Here, $f^{\mathrm{eq}}_i$, $g^{\mathrm{eq}}_i$ and $k^{\mathrm{eq}}_i$ are the local equilibrium distribution functions.
The relaxation parameters $\tau$, $\tau_{\phi}$ and $\tau_{\psi}$ are related to the transport coefficients in the 
hydrodynamic equations, $\eta$, $M_\phi$ and $M_\psi$, through \cite{Briant2004a,Kusumaatmaja2010b}
\begin{eqnarray}
&\eta = \rho c_\mathrm{s}^2 \left(\tau-\frac{\Delta t}{2}\right), \label{ShearViscosity} \\
& M_\phi =  \Gamma_\phi \left(\tau_\phi - \tfrac{\Delta t}{2}\right) \label{Mobility1}, \\
& M_\psi =  \Gamma_\psi \left(\tau_\psi - \tfrac{\Delta t}{2}\right) \label{Mobility2}
\end{eqnarray}
where $\Gamma_\phi$ and $\Gamma_\psi$ are tunable parameters that appear in the equilibrium distribution (see below). 
Since $\eta$, $M_\phi$ and $M_\psi$ are positive quantities, the values of the relaxation times $\tau$, $\tau_\phi$ 
and $\tau_\psi$ have to be larger than $\Delta t/2$. To enforce no slip boundary conditions, we have also implemented 
standard bounce back boundary conditions \cite{Ladd1994b} for the populations of the nodes in contact with the solid boundaries.

Performing a Chapman-Enskog analysis \cite{Chapman1970}, it is possible to show that the LB equations recover the continuity, 
Navier-Stokes and the Cahn-Hilliard equations in the continuum limit if the correct thermodynamic and hydrodynamic 
information is introduced in the simulation by a suitable choice of the local equilibrium functions.
The forms of $f^{\mathrm{eq}}_i$, $g^{\mathrm{eq}}_i$ and $k^{\mathrm{eq}}_i$ that satisfy these requirements for $i > 0$ are 
\cite{Pooley2008b,Kusumaatmaja2010b}
\begin{widetext}
\begin{eqnarray}
 f^{\mathrm{eq}}_i &=& w_i \left(\frac{p_\mathrm{b}}{c_\mathrm{s}^2} + \frac{e_{i \alpha} \rho v_{\alpha}}{c_\mathrm{s}^2} 
 + \frac{\rho v_\alpha v_\beta \left(e_{i\alpha} e_{i\beta} - c_\mathrm{s}^2\delta_{\alpha \beta}\right)}{2 c_\mathrm{s}^4}\right) 
 -\frac{w_i}{c_\mathrm{s}^2}\left(\kappa_{\rho\rho}\rho\nabla^2\rho + \kappa_{\phi\phi}\phi\nabla^2\phi + \kappa_{\psi\psi}\psi\nabla^2\psi \right) \nonumber \\  &&+\frac{\kappa_{\rho\rho}}{c_\mathrm{s}^2}\left(w_i^{xx}\partial_x \rho\partial_x \rho + w_i^{yy}\partial_y \rho\partial_y \rho
 +w_i^{zz}\partial_z \rho\partial_z \rho + w_i^{xy}\partial_x \rho\partial_y \rho  
 +w_i^{yz}\partial_y \rho\partial_z \rho + w_i^{zx}\partial_z \rho\partial_x \rho  \right) \nonumber \\
 &&+\frac{\kappa_{\phi\phi}}{c_\mathrm{s}^2}\left(w_i^{xx}\partial_x \phi\partial_x \phi + w_i^{yy}\partial_y \phi\partial_y \phi
 +w_i^{zz}\partial_z \phi\partial_z \phi + w_i^{xy}\partial_x \phi\partial_y \phi 
 +w_i^{yz}\partial_y \phi\partial_z \phi + w_i^{zx}\partial_z \phi\partial_x \phi  \right) \nonumber \\
 &&+\frac{\kappa_{\psi\psi}}{c_\mathrm{s}^2}\left(w_i^{xx}\partial_x \psi\partial_x \psi + w_i^{yy}\partial_y \psi\partial_y \psi  
 +w_i^{zz}\partial_z \psi\partial_z \psi + w_i^{xy}\partial_x \psi\partial_y \psi 
 +w_i^{yz}\partial_y \psi\partial_z \psi + w_i^{zx}\partial_z \psi\partial_x \psi  \right) \nonumber \\
 &&+\frac{2\kappa_{\rho\phi}}{c_\mathrm{s}^2}\left(w_i^{xx}\partial_x \rho\partial_x \phi + w_i^{yy}\partial_y \rho\partial_y \phi  
 +w_i^{zz}\partial_z \rho\partial_z \phi  \right) -\frac{w_i}{c_\mathrm{s}^2}\left(\kappa_{\rho\phi}\rho\nabla^2\phi + \kappa_{\rho\phi}\phi\nabla^2\rho\right)\nonumber \\
 &&+\frac{\kappa_{\rho\phi}}{c_\mathrm{s}^2}\left(w_i^{xy}\partial_x \rho\partial_y \phi + w_i^{xy}\partial_y \rho\partial_x \phi  
 +w_i^{yz}\partial_y \rho\partial_z \phi + w_i^{yz}\partial_y \rho\partial_z \phi  
 +w_i^{zx}\partial_z \rho\partial_x \phi + w_i^{zx}\partial_z \rho\partial_x \phi \right)   \nonumber \\
 &&+\frac{2\kappa_{\rho\psi}}{c_\mathrm{s}^2}\left(w_i^{xx}\partial_x \rho\partial_x \psi + w_i^{yy}\partial_y \rho\partial_y \psi  
 +w_i^{zz}\partial_z \rho\partial_z \psi  \right) -\frac{w_i}{c_\mathrm{s}^2}\left(\kappa_{\rho\psi}\rho\nabla^2\psi +\kappa_{\rho\psi}\psi\nabla^2\rho\right)\nonumber \\
 &&+\frac{\kappa_{\rho\psi}}{c_\mathrm{s}^2}\left(w_i^{xy}\partial_x \rho\partial_y \psi + w_i^{xy}\partial_y \rho\partial_x \psi  
 +w_i^{yz}\partial_y \rho\partial_z \psi + w_i^{yz}\partial_y \rho\partial_z \psi  
 +w_i^{zx}\partial_z \rho\partial_x \psi + w_i^{zx}\partial_z \rho\partial_x \psi \right)   \nonumber \\
 &&+\frac{2\kappa_{\phi\psi}}{c_\mathrm{s}^2}\left(w_i^{xx}\partial_x \phi\partial_x \psi + w_i^{yy}\partial_y \phi\partial_y \psi  
 +w_i^{zz}\partial_z \phi\partial_z \psi  \right) -\frac{w_i}{c_\mathrm{s}^2}\left(\kappa_{\phi\psi}\phi\nabla^2\psi + \kappa_{\phi\psi}\psi\nabla^2\phi \right)\nonumber \\
 &&+\frac{\kappa_{\phi\psi}}{c_\mathrm{s}^2}\left(w_i^{xy}\partial_x \phi\partial_y \psi + w_i^{xy}\partial_y \phi\partial_x \psi  
 +w_i^{yz}\partial_y \phi\partial_z \psi + w_i^{yz}\partial_y \phi\partial_z \psi  
 +w_i^{zx}\partial_z \phi\partial_x \psi + w_i^{zx}\partial_z \phi\partial_x \psi \right) , \label{eq:eqdistfbinary}  \\
g^{\mathrm{eq}}_i & =&  w_{i} \left( \frac{\Gamma_\phi \mu_\phi}{c_\mathrm{s}^2}  
+ \frac{\phi e_{i \alpha} v_{\alpha}}{c_\mathrm{s}^2} + \frac{\phi v_\alpha v_\beta \left(e_{i\alpha} e_{i\beta} 
- c_\mathrm{s}^2\delta_{\alpha \beta}\right)}{2 c_\mathrm{s}^4}\right),  \\
k^{\mathrm{eq}}_i & =&  w_{i} \left( \frac{\Gamma_\psi \mu_\psi}{c_\mathrm{s}^2} 
+ \frac{\psi e_{i \alpha} v_{\alpha}}{c_\mathrm{s}^2} + \frac{\psi v_\alpha v_\beta \left(e_{i\alpha} e_{i\beta} 
- c_\mathrm{s}^2\delta_{\alpha \beta}\right)}{2 c_\mathrm{s}^4}\right).
\end{eqnarray}
\end{widetext}

For the D3Q19 model, the weights are 
$w_{1-6} = 1/18$, $w_{7-18} = 1/36$, 
$w_{1,2}^{xx} = w_{3,4}^{yy}=w_{5,6}^{zz}=5/36$,
$w_{3-6}^{xx} = w_{1,2,5,6}^{yy}=w_{1-4}^{zz}=-1/9$,
$w_{7-10}^{xx} = w_{15-18}^{xx} = w_{7-14}^{yy}=w_{11-18}^{zz}=-1/72$,
$w_{11-14}^{xx} = w_{15-18}^{yy}=w_{7-10}^{zz}=1/36$,
$w_{1-6}^{xy} = w_{1-6}^{yz}=w_{1-6}^{zx}=0$,
$w_{7,10}^{xy} = w_{11,14}^{yz}=w_{15,18}^{zx}=1/12$,
$w_{8,9}^{xy} = w_{12,13}^{yz}=w_{16,17}^{zx}=-1/12$,
$w_{11-18}^{xy} = w_{7-10}^{yz}= w_{15-18}^{yz}=w_{7-14}^{zx}=0$. 
Furthermore, the speed of sound is $c_\mathrm{s} = (1 / \sqrt{3}) \Delta x / \Delta t$. The equilibrium distribution functions for $i = 0$ can
be computed by ensuring the following relations are satisfied:
\begin{eqnarray}
 & \rho({\bf r}, t) = \sum_i f^{\rm{eq}}_i({\bf r}, t) = \sum_i f_i({\bf r}, t), \nonumber \\
 & \phi({\bf r}, t) = \sum_i g^{\rm{eq}}_i({\bf r}, t) = \sum_i g_i({\bf r}, t), \\
& \psi({\bf r}, t) = \sum_i k^{\rm{eq}}_i({\bf r}, t) = \sum_i k_i({\bf r}, t). \nonumber
\end{eqnarray}

\begin{figure*}
  \centering
  \includegraphics[scale=1.0]{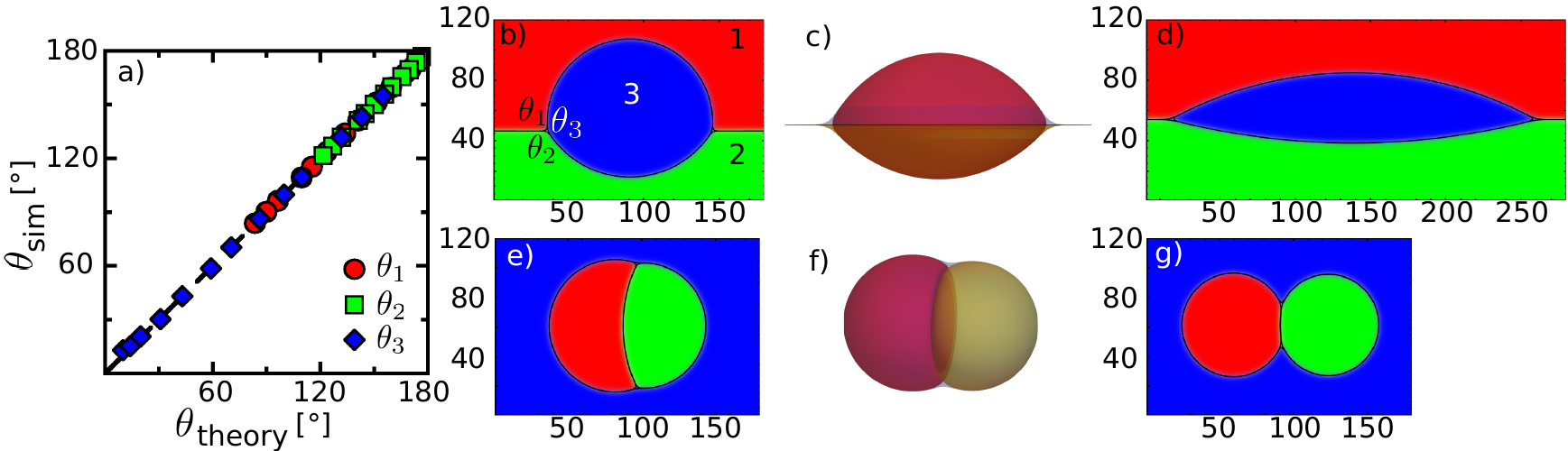}
  \caption{(Color online) a) Comparison between analytical predictions for the Neumann angles 
  $\theta_1$, $\theta_2$ and $\theta_3$ and the simulation results for a liquid lense (panels (b--d)) 
  and a double emulsion (panels (e--g)) at equilibrium. 
  In the simulation, $\alpha=1$, $\kappa_1=0.01$ and $\kappa_2=0.02$ are fixed, while $\kappa_3$ is varied. Specifically, 
  $\kappa_3 =0.05$ in panels (b, e), $\kappa_3=0.15$ in panels (c, f) and $\kappa_3=0.001$ in panels (d, g). We have
  also used $\tau_\phi=1.0$, $\tau_\psi=2/3$ and $\Gamma_\phi = \Gamma_\psi=1.0$.
  Panels (c, f) show the isosurfaces $C=0.5$ for three-dimensional simulation results.
  All remaining panels are for two-dimensional systems.}
  \label{fig:drops}
\end{figure*}

\section{Benchmark Results}
\label{sec:Benchmarks}

We now present a series of systematic benchmarks to show that our model captures the correct equilibrium and 
dynamic behaviors of the ternary fluids. Although our code is capable of handling full 3D geometries, a 2D 
setup is sufficient for the scope of these tests.
In section~\ref{sec:LiquidLens} we start with the liquid lense and double emulsion to test the surface tensions.
We investigate the accuracy of the solid wetting properties in section~\ref{sec:ContactAngles}.
Finally, in section~\ref{sec:PhaseSeparation}, we examine different scenarios for ternary phase separation.

\subsection{Liquid Lense and Double Emulsion}
\label{sec:LiquidLens}

The first set of simulations are designed to verify the fluid-fluid surface tensions against the 
analytical predictions in Eq.~\eqref{eq:SurfTens}.
To do this, we have simulated liquid lenses (panels (b)--(d)) and double emulsions 
(panels (e)--(g)), as shown in Fig.~\ref{fig:drops}.
In both cases, the force balance between all three surface tensions must be satisfied at the contact line, 
which we can succinctly write in vector notation:
\begin{equation}
\vec{\gamma}_{12} + \vec{\gamma}_{23} + \vec{\gamma}_{31} = \vec{0}.
\end{equation}
Due to this force balance, the three angles $\theta_1$, $\theta_2$ and $\theta_3$ in 
Fig.~\ref{fig:drops} satisfy the \emph{Neumann triangle relation}
\begin{equation}
\frac{\gamma_{12}}{\sin \theta_3} = \frac{\gamma_{23}}{\sin \theta_1} = \frac{\gamma_{31}}{\sin \theta_2}.
\end{equation}

In Fig.~\ref{fig:drops}(a), we systematically compare the angles $\theta_1$, $\theta_2$ and $\theta_3$ 
obtained from the simulations and analytical predictions for the liquid lense geometry.
The simulation box has a height of $120$ l.u.~(lattice units) and a width ranging from $120$ l.u.~to $260$ l.u.~in order to
accommodate the lense geometry.
To ensure the systems have reached mechanical equilibrium, we typically run the simulations until 
the maximum fluid velocity in the whole simulation domain is less than a given threshold value. 
Here, we have used $10^{-5} \Delta{}x / \Delta{}t$, which corresponds to the maximum spurious velocity 
in our simulation, close to the fluid-fluid interface. This is consistent with usual results in LB literature.

As a representative example, we choose $\kappa_2/\kappa_1 = 2$, and 
incrementally vary $\kappa_3$ over a large range of parameters. 
The detailed choice of simulation parameters is reported in the caption of Fig.~\ref{fig:drops}.
Note that the surface tension is $\gamma_{mn} = \alpha (\kappa_m + \kappa_n)/6$ 
in our model. As shown in the figure, our simulations accurately reproduce the predicted angles. 
We only observe significant variations between the measured and analytical values ($> 3^\circ$) when one
of the Neumann angles is less than $10^\circ$ ($\theta_3$ in the case shown in Fig.~\ref{fig:drops}). 
This is caused by the diffuse nature of the fluid-fluid interfaces. 
We obtain the same level of accuracy when the test is carried out for the double emulsions (Fig.~\ref{fig:drops}(e--g)).

Additionally, we have carried out the Laplace pressure test for a droplet of component $m$ surrounded by 
fluid $n$ (data not shown). We performed all possible pairwise permutations to ensure that our model is 
still symmetric after the introduction of the auxiliary variables. In such a case, the simulation basically 
reduces to a binary fluid system; and indeed the results are identical to those from a binary free energy 
model, as expected. 

\subsection{Contact Angles}
\label{sec:ContactAngles}

\begin{figure}
  \centering
  \includegraphics[scale=1.0]{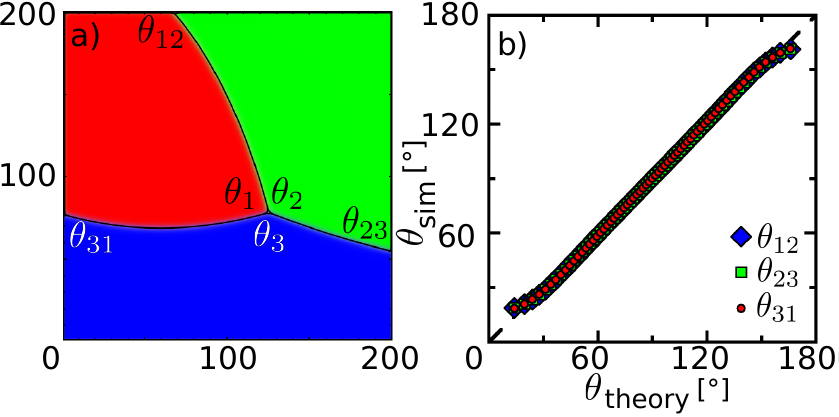}
  \caption{(Color online) 
  Validation of the wetting boundary conditions at solid walls: simulations in both panels 
  are carried out with $\alpha=1$, $\tau_\rho = 1.0$, $\tau_\phi=1.0$, $\tau_\psi=2/3$ and
  $\Gamma_\phi = \Gamma_\psi=1.0$.
  a) Three fluid phases at equilibrium confined to a square well, showing simultaneously the 
  three Neumann angles and the three different pairs of contact angles.
  Here $\kappa_1=0.01$, $\kappa_2=0.02$, $\kappa_3=0.03$, $h_1=-0.002$, $h_2=0.002$ and $h_3=0.0$.
  b) Comparison between the  predicted and simulated contact angles as the parameters $h_1$, $h_2$, and 
  $h_3$ are varied. Here $\kappa_1=\kappa_2=\kappa_3=0.01$,  The agreement departs for very small and very 
  large angles. The minimum and maximum values shown are $\theta = 14^\circ$ and $\theta = 166^\circ$, respectively.
  Outside this range, the deviation in contact angle is $>4^\circ$ between prediction and measurement.}
  \label{fig:angles}
\end{figure}

The next set of verifications concerns with the contact angles of the fluids at a solid surface.
We have introduced the geometry shown in Fig.~\ref{fig:angles}(a) where the ternary fluid
is confined to a square well. Such a setup  demonstrates that we can simultaneously control
all three contact angles given the parameters $h_1$, $h_2$, and $h_3$, including the Girifalco-Good
relation in Eq.~\eqref{eq:GoodGirifalco}. Additionally, we can also recover the Neumann triangle at 
the point where the three fluid-fluid interfaces meet, as discussed in section~\ref{sec:LiquidLens}.

Fig. \ref{fig:angles}(b) shows the measured contact angles in our simulations compared to the 
analytical predictions, see Eq.~\eqref{eq:SurfTens} and Eq.~\eqref{eq:solidST}. 
The simulation parameters are reported in the figure caption.
The contact angle is given by Young's formula in Eq.~\eqref{eq:contact angle}. 
Similar to the Neumann angles, the simulation results for the contact angles are accurate with deviations 
of up to $4^\circ$ at contact angles $14^\circ$ and $166^\circ$. This discrepancy is again due to the finite 
width of the interface. For very small or large contact angles, the diffuse fluid-fluid and fluid-solid 
interface profiles interfere with one another and affect the contact angle results. 
The same issue is observed for the binary fluid model \cite{Pooley2008}.

\subsection{Ternary Phase Separation}
\label{sec:PhaseSeparation}
  
\begin{figure*}
  \centering
  \includegraphics[scale=1.0]{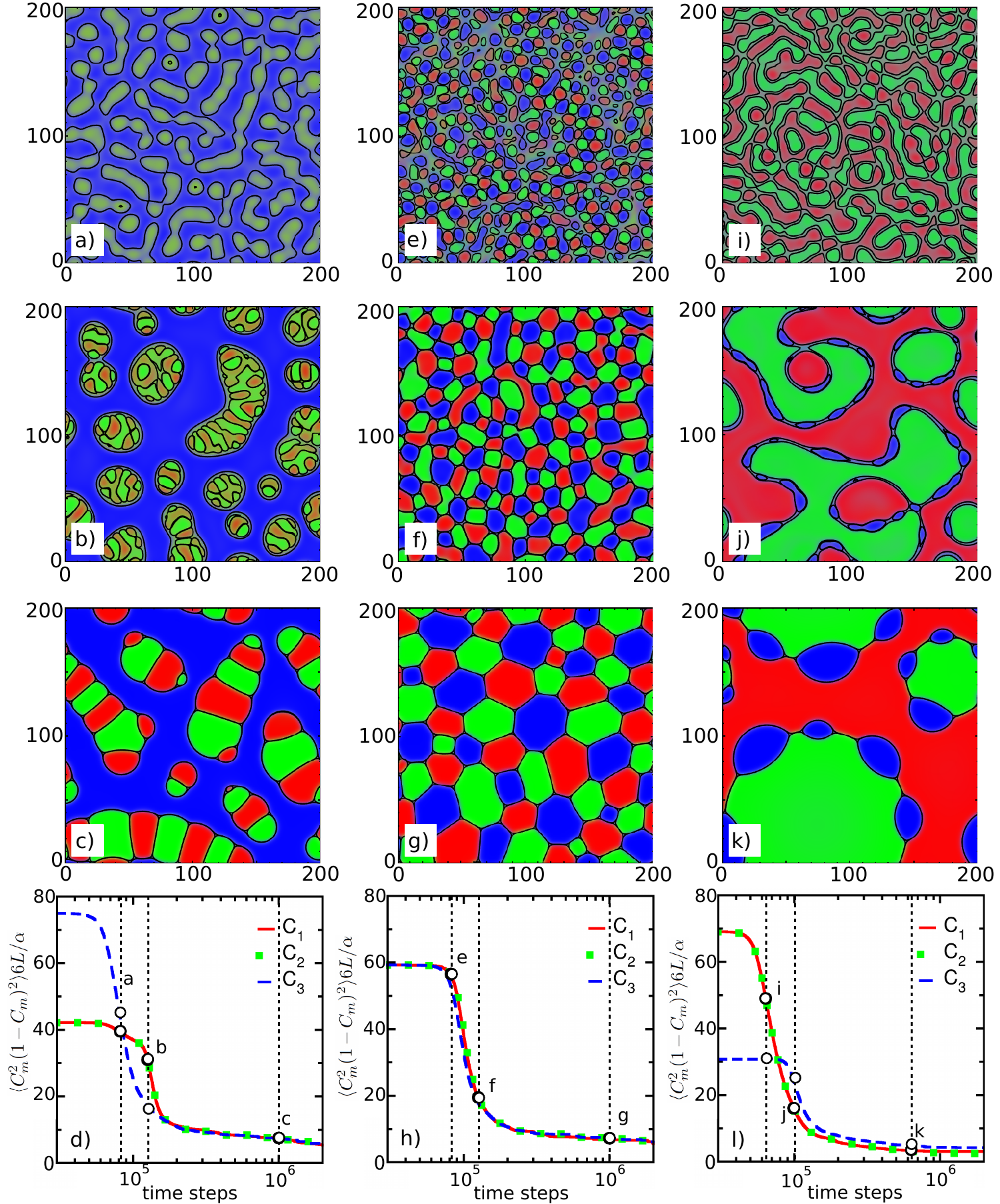}
  \caption{(Color online) Ternary phase separation for three different sets of fluid compositions: (i) $C_1=C_2=1/4$ 
  and $C_3=1/2$ (panels (a--d)), (ii) $C_1=C_2=C_3=1/3$ (panels (e--h)), and (iii) $C_1=C_2=2/5$ and 
  $C_3=1/5$ (panels (i--l)). Here we use  $\kappa_1=\kappa_2=\kappa_3=0.01$, $\tau_\rho = 1.0$, $\tau_\phi=1.0$, 
  $\tau_\psi=2/3$, $\Gamma_\phi = \Gamma_\psi=1.0$ and the fluid compositions are initialized with random perturbations of amplitude $\delta = 0.01$. 
  Panels (d), (h) and (l) show the time evolution of the quantity $\chi_m=\langle C_m^2(1-C_m)^2  \rangle 6L/\alpha$, 
  where $L=200 \Delta x$ is the size of the square domain in lattice units and $m = (1,2,3)$. The rapid decay 
  of $\chi_m$ indicates the segregation of the phase $C_m$. Once the phase is well separated, its value 
  represents the total length of the boundary between $C_m$ and the other two phases. The black circles and 
  vertical dashed lines indicate the time of snapshots reproduced in the upper panels.}
  \label{fig:phase_separation}
\end{figure*}

So far we have verified the equilibrium thermodynamics of our ternary fluid model.
Here we will shift our attention to the dynamic behavior of the fluids, by studying the phase 
separation of ternary fluids and how this depends on the composition of the fluids, $(C_1, C_2, C_3)$. 
We use the following three compositions: (i) $(0.15, 0.15, 0.7)$, (ii) $(1/3, 1/3, 1/3)$, and (iii) 
$(0.4, 0.4, 0.2)$ to allow comparisons with previously published results on this topic \cite{chen_computer_1994}.
It is worth noting that the results in \cite{chen_computer_1994,Blowey1996} are obtained by solving the Cahn-Hilliard 
equations with zero fluid velocity. Here we simulate the fully coupled thermodynamic-hydrodynamic system.
The simulation parameters are reported in the figure caption of Fig. \ref{fig:phase_separation}.

In all cases we initialize the simulations by introducing random concentration fluctuations with 
amplitude $\delta = 0.01$ to an otherwise homogeneously mixed fluid. Since the system is unstable 
with respect to concentration fluctuations, spinodal decomposition then takes place; the system 
separates into spatial regions rich in one phase and poor in the other phases. Such a process 
reduces the overall free energy of the system.

The dynamics of the spinodal decomposition depends strongly on the composition. In case (i), where $C_3$
is dominant and $C_1$ and $C_2$ are minorities, the system initially separates into domains of $C_3$ and a 
mixture between $C_1$ and $C_2$ (see Fig.~\ref{fig:phase_separation}(a)). It is only at later times that 
$C_1$ and $C_2$ themselves phase separate (panels (b) and (c)). Interestingly, we observe a ``worm-like'' 
structure where domains of $C_1$ and $C_2$ form layers and together they are encapsulated by $C_3$. 
Further coarsening occurs due to rearrangements and subsequent coalescence of neighbouring $C_1$ and $C_2$ 
domains.

Let us now consider the symmetric case where all three fluids are equally represented, case (ii). Here,
as expected, the coarsening dynamics is equivalent for all three fluids. The fluids initially
form small droplets (Fig.~\ref{fig:phase_separation}(e)) that then grow due to a combination of diffusion 
and coalescence. In panels (f) and (g), we report the ternary network of of domains, whose typical size
coarsen at the same speed. 

The situation is very different when two fluid components are dominant and the third fluid is a minority, 
case (iii). As shown in Fig.~\ref{fig:phase_separation}(i), the system initially behaves akin to a binary 
system. $C_1$ and $C_2$ phase separate, with the fluid component $C_3$ trapped at the interfaces between
$C_1$ and $C_2$. Only when the majority fluid phases have coarsened considerably, the minority fluid
starts to show a bulging effect where droplets of $C_3$ form at the interfaces between $C_1$ and $C_2$,
see panels (j) and (k).

We can quantitatively trace the different onsets of phase separation  by plotting the average quantity 
$\chi_m=\langle C_m^2(1-C_m)^2 \rangle 6L/\alpha$ as a function of time. $\chi_m=0$ when fluid $m$ takes either 
$C_m = 0$ or $1$, and $\chi_m>0$ otherwise. As shown in Fig.~\ref{fig:phase_separation}(d),(h) and (l), 
the rapid decay of $\chi_m$ marks the onset of phase separation of fluid component $m$.
Furthermore, after the system has phase separated, $\chi_m$ provides an estimate for the total length
of the boundary of the component $m$. In case (i), we see that  $\chi_3$ decays first, followed by 
$\chi_1$ and $\chi_2$. The opposite is observed for case (iii), while in case (ii), all three quantities decay 
at the same time.

We note that the sequence of morphologies reported here as the fluids undergo phase separation and how they depend
on the fluid concentrations are qualitatively consistent with the results in \cite{chen_computer_1994,Blowey1996}. 
Quantitative comparisons, however, are not feasible because these works do not account for hydrodynamics
and they have not reported detailed time sequence data.

\begin{figure}
  \centering
  \includegraphics[width=1.0\columnwidth]{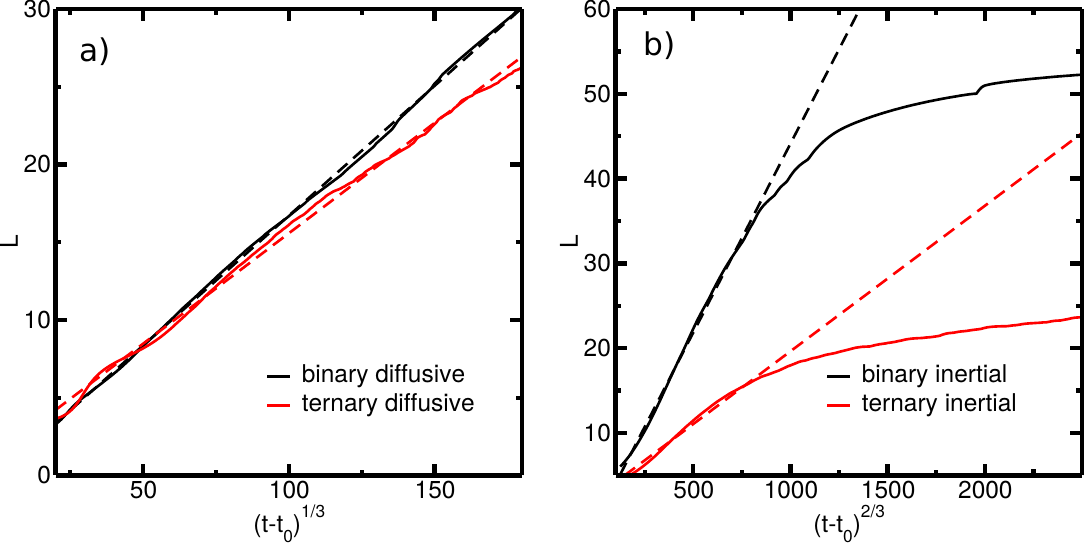}
  \caption{
  (Color online) Scaling of the typical domain size $L=A/\chi$, during phase separation in 
  two dimensions. a) Diffuse regime.  b) Inertial regime.
  The area of the domain size is $400\times400$ lattice spacings. Ternary and binary mixtures 
  are introduced by setting $C_1=C_2=C_3=1/3$ and $C_1=C_2=1/2$ respectively.  The parameters 
  $\tau_\rho = 1.0$, $\tau_\phi=1.0$, $\tau_\psi=2/3$, $\Gamma_\phi = \Gamma_\psi=1.0$ are the 
  same in all cases. For the diffuse regime we set $\kappa_1=\kappa_2=\kappa_3=0.01$, while
  for the inertial regime we set $\kappa_1=\kappa_2=\kappa_3=0.04$.
  Each curve is shifted by its time constant $t_0$ obtained from a fitting routine.
  }
  \label{fig:scaling}
\end{figure}

To test the dynamic behaviour of the system we compare the evolution of the typical domain size,
estimated by $L=A/\chi$ where $A$ is the area of the periodic domain. We simulate both binary and ternary 
mixtures with symmetric concentration of phases. For the diffusive regime, obtained by removing the 
advection term in the Cahn-Hilliard equation,
we recover the theoretical scaling of $L\sim t^{1/3}$. For the inertial regime the theoretical scaling
$L\sim t^{2/3}$ is observed only at the onset of phase separation. While in the binary case the coarsening 
stops due to finite size effects, in the ternary case the coarsening is inhibited at an earlier stage by the 
formation of a foam-like network as depicted in Fig. \ref{fig:phase_separation}g. After this structure has formed, 
the only mechanism leading to further coarsening is related to Ostwald ripening, characterised by slow 
diffusion followed by sporadic coalescence events that change the network topology. We argue that the situation would 
be different in three dimensions, as at least one additional phase is required to establish a stable packing. 
We plan to report in details the phenomenon of the ternary phase separation in a forthcoming publication.

\section{Conclusions}
\label{sec:conclusion}

We have presented a ternary free energy lattice Boltzmann (LB) model. In section~\ref{sec:ThermoDynamics} 
we analytically derived the values of the fluid-fluid surface tensions, the fluid-solid surface tensions 
and the contact angles, given the simulation parameters $\kappa_m$ and $h_m$ ($m = 1, 2, 3$). The three 
$\kappa$ parameters can be varied independently, thus allowing us to arbitrarily tune the three fluid-fluid 
surface tensions. Due to the Girifalco-Good relation, only two out of the three solid surface contact angles 
are actually independent. Here, we not only show how the contact angles can be computed given the $h_m$, 
but we also describe the procedure to invert the contact angle relations. Our free energy formulation also 
allows additional fluid components to be added easily if required by an application. 

The free energy LB model is a top-down approach. Given the free energy functional, we can subsequently 
derive the chemical potentials, pressure tensor, and wetting boundary conditions (section~\ref{sec:EoM}) 
that need to be introduced in the LB algorithm (section~\ref{sec:LBM}). We have also shown in section~\ref{sec:Benchmarks} 
that our model and algorithm are able to capture the correct equilibrium and dynamic behavior of the 
ternary fluids. In particular, we considered three benchmark tests: i) double emulsion and liquid lense, 
(ii) ternary fluids in contact with a square well, and (iii) ternary phase separation.

Though it is beyond the scope of this paper, we note that there is a wide range of applications that 
our model can simulate. For example, we can study the dynamics of ternary emulsions
 \cite{Utada2005,Park2012,Guzowski2015,Haase2014} and compound 
droplets \cite{gao_spreading_2011, luo_deformation_2015}. The ternary phase separation, 
as shown in section \ref{sec:Benchmarks}, is also an extremely rich phenomenon that is worth being studied in 
more detail. Furthermore, topographical and/or chemical heterogeneities can be accounted easily in LB simulations. 
This is important, e.g., for simulations of liquid infused surfaces 
\cite{Wong2011,Smith2013,Schellenberger2015, wexler_shear-driven_2015} that involve a 
ternary fluid system (water, oil and air) in conjunction with rough surfaces.

Finally, we note that in the current model all the fluid densities are equal and set to $n = 1$. 
This is a reasonable approximation for when all three fluids have similar densities, or when the 
Reynolds number in the problem of interest is small. To tackle phenomena where inertial terms are 
relevant, another important area of future work is to generalise our model to allow different densities. 
In the context of the ternary free energy lattice Boltzmann model introduced here, this corresponds to 
devising a free energy functional with minima located at different values of $n$, not just $\phi$ and $\psi$.

\section*{Acknowledgements}

CS and HK thank Muhammad Subkhi Sadullah, Dr Matthew Wagner, Dr Yonas Gizaw and Dr Peter Koenig for fruitful 
discussions, and acknowledge Procter and Gamble (P\&G) for funding.
TK thanks the University of Edinburgh for the award of a Chancellor's Fellowship.

\appendix

\section{Mobility Parameters}
\label{app:MobilityParameters}

Here we will derive the limiting case where all fluid phases have symmetric mobility parameters,
$M_1 = M_2 = M_3 = M$, and show that this corresponds to $M_\phi = 3 M_\psi$ in our model. 
We start from the Cahn-Hilliard equation for each fluid component $m$, given by
\begin{equation} 
        \partial_t C_m  + \partial_\alpha (C_m v_\alpha)  = M_m \nabla^2 \mu_{m} . \label{eq:CH3}
\end{equation}
The key point to realize is that the three fluid components are not independent
since they are related by the constraint $C_1 + C_2 + C_3 = 1$. In Eq.~\eqref{eq:CH3}, this can 
be handled by introducing a Lagrange multiplier $\beta$ such that \cite{Boyer2006}
\begin{equation}
\mu_m = \frac{\partial F}{\partial C_m} + \beta , \quad \mu_1 + \mu_2 + \mu_3 = 0.
\end{equation}
Substituting these relations into Eq.~\eqref{eq:CH3} and comparing the results with Eq.~\eqref{eq:Cahn-Hilliard1}
and Eq.~\eqref{eq:Cahn-Hilliard2}, it can be shown that we have $M = M_\phi/2$ and $M = 3M_\psi/2$
for the limiting case of symmetric mobility parameters. Correspondingly, we thus have $M_\phi = 3 M_\psi$.

\bibliography{references}

\end{document}